\documentclass[aps, twocolumn,prb,showpacs]{revtex4}

\usepackage{graphicx}
\usepackage{dcolumn}
\usepackage{bm}

\begin{document}

\title{Dynamic nuclear polarization and spin-diffusion in non-conducting solids}\

\author{Chandrasekhar Ramanathan\footnote{Electronic address:
sekhar@mit.edu}}
\affiliation{Department of Nuclear Science and Engineering, Massachusetts Institute of Technology, Cambridge, MA 02139, USA}

\date{\today}

\begin{abstract}
There has been much renewed interest in dynamic nuclear polarization (DNP), particularly in the context of solid state biomolecular NMR and more recently dissolution DNP techniques for liquids. This paper reviews the role of spin diffusion in polarizing nuclear spins and discusses the role of the spin diffusion barrier, before going on to discuss some recent results.

\end{abstract}

\maketitle

Microwave irradiation of a coupled electron-nuclear spin system can facilitate a transfer of polarization from the electron to the nuclear spin.  In dielectric materials dynamic nuclear polarization (DNP) typically occurs via the solid effect,thermal mixing or the cross effect \cite{Atsarkin-1978,Abragam-1982,Reynhardt-2003b}.  In these systems the electron spins are localized and the non-equilibrium polarization of the bulk nuclei is generated via a two-stage process:  a polarization exchange local to the defect; and spin transport to distribute the polarization throughout the sample.   Here the DNP process is essentially the inverse of the standard $T_1$ relaxation mechanism in dielectric solids \cite{Bloembergen-1949}.  

While much attention is paid to improving the local polarization transfer efficiency from the electron to the neighboring nuclear spins, usually by incorporating the appropriate electron spins in the sample, the rate-limiting step in efficiently polarizing bulk samples is frequently the spin transport from the defect sites to the bulk.  
As we attempt to use DNP to enhance nuclear magnetization in an ever-increasing number of systems, it is important to understand the many-spin dynamics that underly the process.  As we improve our system model for describing DNP dynamics, we will eventually be able to incorporate recent developments in the theory and practice of optimal control of quantum systems to further improve our techniques\cite{Yao-1991,Khaneja-2001}.  

The purpose of this paper is to clarify our understanding of the spin dynamics in light of recent experiments in our laboratory.  The paper begins with a description of the Bloembergen model and examines the different steps of the transport processes involved in dynamic nuclear polarization, before concluding with a review of recent experimental results.

In a strong magnetic field, the Hamiltonian of a nuclear spin system, doped with electron spins is given by
\begin{equation}
\mathcal{H}_{\mathrm{tot}} = \mathcal{H}_Z^n + \mathcal{H}_Z^{e} + \mathcal{H}_D^n + \mathcal{H}_D^e + \mathcal{H}_E^e + \mathcal{H}_{HF}
\end{equation}
corresponding to the nuclear and electron Zeeman interactions, the nuclear-nuclear and electron-electron dipolar interactions, the electron exchange interaction and the electron-nuclear hyperfine interaction (including both the Fermi contact and the electron-nuclear dipolar 
interactions).  During the DNP process we could add both microwave and RF fields to irradiate the electronic and nuclear spins as needed. The challenge of dealing with this full Hamiltonian in a systematic quantum mechanical formalism underlies the gaps in our knowledge of the full DNP process.  Traditionally researchers have adopted a composite quantum-classical approach in which an isolated defect spin coupled to one or two nuclear spins is treated quantum-mechanically, and a classical approach is used to deal with the transport of the polarization either to or away from the defect.

\subsubsection*{The Bloembergen model}
DNP and spin-lattice relaxation are very similar processes in these systems.
Waller \cite{Waller-1932} provided the first theoretical treatment of nuclear magnetic relaxation in ionic crystals, in which the spin-spin interactions between nuclei were modulated by lattice vibrations. However observed relaxation times were much shorter than those calculated with this model.  In 1949 Bloembergen \cite{Bloembergen-1949} postulated that NMR relaxation in these non-conducting solids was mediated by nuclear spin diffusion from the bulk to the sites of paramagnetic impurities, and that the T$_1$ were significantly shortened at increased doping densitites.

The dynamics of the nuclear polarization ($p$) can therefore be described by a diffusion equation in the continuum limit,
\begin{equation}
-\frac{\partial p}{\partial t}  =  D \nabla^2 p  + 2Wp + \frac{C}{r^6}\left(p-p_0\right)\label{eq:Bloem}
\end{equation}
where $D$ is the diffusion coeffient, $W \approx \gamma^2 B_1^2 T_2$ is the rate at which the applied RF drives nuclear spin transitions, and $C/r^6$ describes the rate at which the nuclear spins at a distance $r$ from the impurity are relaxed.  Here, a single defect spin is surrounded by a large number of nuclei, in the presence of an applied RF field.  The equation can be summed over all the defect spins to describe the dynamics of the entire sample.

\begin{figure}
\scalebox{0.4}{\includegraphics{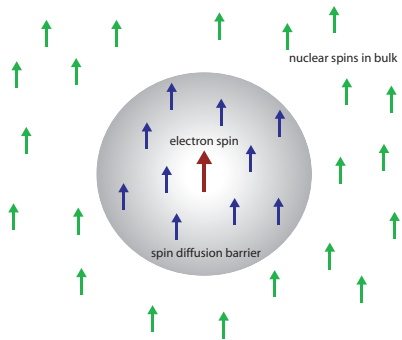}} \caption{\label{diffusionbarrier} Schematic illustration of the spin diffusion barrier around the electron spin.  Nuclear spin diffusion within the barrier is suppressed due to the large difference in Zeeman energies between the spins. (from Ref.(8))} 
\end{figure}

The spin diffusion process is mediated by energy-conserving flip-flop transitions that take place during evolution under the dipolar Hamiltonian.  In a strong external magnetic field, the secular dipolar Hamiltonian can be written as
\begin{equation}
\mathcal{H}_D = \sum_{i,j} d_{ij} \left( 2I_z^iI_z^j - \frac{1}{2}\left(I_+^iI_-^j + I_-^iI_+^j\right)\right)
\end{equation}
where $d_{ij} = \gamma^2\hbar^2 (1-3\cos^2\theta_{ij})/2r_{ij}^3$, $r_{ij}$ is the distance between spins $i$ and $j$, and $\theta_{ij}$ is the angle between the internuclear vector and the external magnetic field.
However, the nuclear spins in an inner core around the impurity experience a large local field gradient due to the impurity spin, and as a consequence have significantly different effective Zeeman energies (Figure \ref{diffusionbarrier}).  This energy difference suppresses the flip-flop terms in the above equation, creating a ``spin-diffusion barrier" around the impurity, within which the polarization is ``frozen'', and the diffusion coeffient is zero. 

Thus in this model, the nuclear spins inside the barrier were relaxed directly by the impurity electron spin, and were isolated from the nuclear spins in the bulk.  At the edge of the barrier, spins are in contact with both the impurity spin as well as the neighboring nuclear spins while the bulk nuclear spins do not experience the field of the impurity spin directly.

\subsubsection*{Relaxation and DNP}
The process of microwave irradiation of an electron spin coupled to a nuclear spin has been well described \cite{Jeffries-1963,Abragam-1982}, and detailed quantum-mechanical treatments of the isolated two spin system are also available \cite{Jeschke-1996,Jeschke-1999,Weis-2000,Hu-2006}.  As the focus of this paper is on the transport processes, we will not discuss this aspect of DNP further.

It was recognized early on that the relaxation process described above was essentially the same physical model that was needed to describe the development of large bulk nuclear spin polarizations following DNP.  Leifson and Jeffries \cite{Leifson-1961} and Khutsishvili \cite{Khutsishvili-1963} modified Bloembergen's equation to incorporate the effect of driving the electron spin transitions.
In the presence of both RF and microwave irradiation of the nuclear and electron spins respectively Khutsishvili \cite{Khutsishvili-1966} obtained the following differential equation for the bulk nuclear spin magnetization ($M$) 
\begin{eqnarray}
\frac{\partial M}{\partial t} & = & \frac{M_0 -M}{T_d} + D\nabla^2 M - C\sum_m \frac{(M - M_0)}{|r-r_m|^6}  \nonumber \\ & & \hspace*{0.3in} - 2WM - \Gamma_{\pm}\sum_m\frac{M \mp \eta M_0}{|r - r_m|^6}
\end{eqnarray}
where $\eta = \gamma_e / \gamma_n$ is the DNP enhancement factor, $T_d$ is the nuclear relaxation rate due to extraneous impurities that do not contribute to DNP, and $\Gamma_{\pm}$ is the DNP driving rate and the sign indicates which ESR transition was being irradiated.  This result does not take into account electron-electron couplings and is thus valid in the limit of a low concentration of paramagnetic impurities.  The boundary conditions for this macroscopic transport equation are determined by the physics in the vicinity of the electron spin, and it is useful to explore this region in more detail.

\subsubsection*{The spin diffusion barrier}

Khutsishvili provided a first formal theory of the spin diffusion barrier \cite{Khutsishvili-1962}.  He defined the barrier to be the distance $d$ from the paramagnetic impurity at which the difference of the hyperfine-shifted Zeeman frequencies of two neighboring nuclei is equal to the nuclear resonance linewidth (dipolar broadened).  Blumberg \cite{Blumberg-1960} defined the barrier to be the distance at which the field due to the ion equals the local dipolar field which results in a slightly larger distance for the barrier. The descriptions were formalized further by Rorschach \cite{Rorschach-1964}
 and Khutsishvili \cite{Khutsishvili-1969} yielding

\begin{equation}
d \approx \left(\frac{2S\gamma_e}{\gamma_n}\right)^{\alpha} a 
\end{equation}
if $\tau > T_2^n$ or  $S\hbar \gamma_e B_0 / kT > 1 $, and
\begin{equation}
d \approx  \left[2S\frac{\gamma_e}{\gamma_n}B\left(\frac{S\hbar\gamma_eB_0}{kT}\right)\right]^{\alpha} a 
\end{equation}
if $\tau < T_2^n$  or $S\hbar \gamma_e B_0/kT < 1$,
where $a$ is the inter-nuclear distance, $S$ is the electron spin, $B(\cdot)$ is the Brillouin function and $\alpha = 1/4$ for Khutsishvili's original definition of the barrier and $\alpha= 1/3$ for the Blumberg definition.  Here $\tau$ is the correlation time of the $S_z$ component of the electron spin.  For dilute spins it corresponds to $T_1^e$, while for dipolar- or exchange-coupled electron spins it corresponds to $T_2^e$.  If the electron spin is fluctuating rapidly on the timescale of the nuclear $T_2$, it is only the mean thermal polarization that needs to be taken into account.  The smaller barrier at lower electron polarization makes it easier to achieve higher DNP enhancments.  However, if the goal is to achieve large nuclear spin polarizations it is necessary to start with a highly polarized electron spin system which results in a large spin diffusion barrier.
It should be noted however that these classically defined models are defined in the continuum limit and assume no anisotropy of the spin diffusion barrier.  
If the barrier is defined by electron-nuclear dipolar interactions, the angular dependence will have the same $(1-3\cos^2\theta)$ dependence with respect to the external field.  

Khutsishvili \cite{Khutsishvili-1957} and de Gennes \cite{deGennes-1958} introduced the pseudopotential radius $b$, also called the scattering parameter.  This distance characterizes the competition between direct relaxation due to the paramagnetic impurity and spin diffusion.  If the distance between impurities is larger than $b$ the T$_1$ relaxation (and DNP) is diffusion-limited, while if the distance between impurites is smaller than $b$, spin diffusion is relatively unimportant in T$_1$ and DNP.  If $d < b$ then the relaxation is limited by the diffusion of the magnetization to the sites of the impurities, while if $d > b$ polarization diffuses to the site of the impurity faster than the paramagnetic impurity can transmit it to the lattice.

Using the definition given by Blumberg ($\alpha = 1/3$), Goldman \cite{Goldman-1965} estimated the radius of the spin diffusion barrier in paradibromobenzene to be 17 \AA.  His measurements suggested a steep decrease of the diffusion coefficient at the spin diffusion barrier.  Schmugge and Jeffries \cite{Schmugge-1965} estimated the size of the barrier in Nd-doped Lanthanum Magnesium Nitrate (LaMN) to be 16 \AA, based on the same model.

However, experiments by a number of other authors suggested that the effective barrier was infact much smaller than this.  Ramakrishna and Robinson \cite{Ramakrishna-1966} were able to study the dynamics of the protons close to the defect site, by first irradiating one forbidden transition for a long time, and then switching the irradiation frequency to the other forbidden transition.   Their results suggested a spin diffusion barrier of 5-7 \AA, which is even smaller than that obtained using  the Khutsishvili definition ($\alpha = 1/4$) which gives 9 \AA. Ramakrishna \cite{Ramakrishna-1966b} was also able to observe a small anisotropy of the barrier.  Tse and Lowe also found the experimentally observed spin diffusion barrier in calcium fluoride was about a factor or 2 smaller than that predicted by the Khutsishvili theory \cite{Tse-1968}.

Using high sensitivity NMR techniques to directly detect the near-nuclei around the paramagnetic impurity, Wolfe and collaborators were able to directly probe the thermal contact between these hyperfine-shifted nuclei and the bulk nuclear spins.  In Yb/Nd doped yttrium ethyl sulphate (YES) they observed that very few spins were not in thermal contact with the bulk, and that the diffusion barrier only contained 1-2 shells of nuclear spins around the impurity, indicating a barrier on the order of 3 \AA \cite{King-1972,Wolfe-1973}.   In a similar experiment on Eu-doped calcium fluoride they observed that only the first shell of nuclear spins was isolated from the bulk \cite{Hansen-1978}.

\subsubsection*{Transport through the spin diffusion barrier}

The discrepancy arises from the assumption that spin diffusion is completely quenched within the barrier.  Polarization transport within the barrier would re-introduce a measure of thermal contact between these nuclei and the bulk.  However, as noted by Bloembergen, transport through the spin diffusion barrier does not conserve Zeeman energy.  Thus it is necessary for this additional energy to be provided by another energy reservoir.

Horvitz \cite{Horvitz-1971} suggested that the fluctuating fields of the electron spin itself could facilitate transport through the barrier.  The electron spin-phonon coupling gives rise to a fluctuating electron spin, which has components at all frequencies, including at the mismatch of the nuclear Zeeman energies. As long as the T$_1$ of the electrons is not too long, this can facilitate transport through the spin diffusion barrier.  Wolfe experimentally observed this effect in both Yb-doped YES and CaF$_2$ \cite{King-1972,Wolfe-1973,Hansen-1978}.  Wolfe \cite{Wolfe-1973} also observed that the electron spin dipolar couplings can facilitate transport through the barrier.  In his experiments he showed that the effective spin diffusion barrier decreased as the concentration of impurities increased.

Goldman \cite{Goldman-1965} suggested that the nuclear dipolar reservoir could make up the energy mismatch.  In a slightly different context Redfield and Yu \cite{Redfield-1968} considered spin diffusion in a macroscopically inhomogeneous field, and noted that this results in a transfer of energy between the nuclear spin Zeeman and dipolar reservoirs.  Genack and Redfield \cite{Genack-1973} observed the coupling of nuclear spin Zeeman and dipolar energy reservoirs in superconducting vanadium. They derived a set of coupled differential equations \cite{Genack-1975} to describe the macroscopic transfer of energy between the Zeeman and dipolar reservoirs.  Neglecting relaxation, they obtained
\begin{equation}
\frac{\partial\beta_d}{\partial t}  =  D_D\nabla^2\beta_d+ \frac{D_Z (\nabla B)}{B_{\textrm{loc}}^2} \cdot \left(\nabla(B\beta_Z) - \beta_D\nabla B\right)
\label{eq:GR-dipolar}
\end{equation}
and
\begin{equation}
\frac{\partial\beta_Z}{\partial t}  =  \frac{D_Z}{B} \nabla \left[\nabla (B\beta_Z) - \beta_d\nabla B \right]
\label{eq:GR-Zeeman}
\end{equation}
where $\beta_d$ and $\beta_{Z}$ are the inverse spin temperatures of the dipolar and Zeeman reservoirs, $D_D$ and $D_Z$ are the spin diffusion rates of dipolar and Zeeman order, $B_{\textrm{loc}}$ is the strength of the local dipolar field and $\nabla B$ is the gradient of the magnetic field around the impurity.  In many samples it is this coupling of the Zeeman and dipolar reservoirs at the impurity or defect sites that permits the transport of polarization through the barrier. This process has recently been re-examined by Furman and Goren \cite{Furman-2003}.
It has also been suggested that if spin diffusion is rapid within the field gradients of the paramagnetic impurity, then the heat capacity of the nuclear spin dipolar reservoir is significantly enhanced, as the hyperfine shifted nuclear spins in the barrier are more tightly coupled to the nuclear dipolar reservoir than the nuclear Zeeman reservoir \cite{Cox-1977}. 

Furman and Goren \cite{Furman-2002} have suggested that it should be possible to short-circuit the spin diffusion barrier, by performing a standard Hartmann-Hahn \cite{Hartmann-1962} cross polarization experiment between the spins within the barrier and those in the bulk.  This would also permit the indirect detection of the spins within the barrier.  It is also possible that the size of the spin diffusion barrier can change during microwave irradiation of the electron spins \cite{Schmugge-1965}.  Under strong microwave irradiation, the electrons can effectively be decoupled from the distant nuclear spins, thus reducing the size of the barrier significantly. 

The presence of the nuclear spin diffusion barrier has also been observed in a variety of ESR experiments.  The dipolar coupling between the electron spin and the more distant nuclear spins gives rise to weak satellite lines in the ESR spectrum \cite{Trammel-1958}.  These so-called ``spin-flip" ($s$) transitions saturate more slowly under microwave irradiation compared to the main ESR line \cite{Sagstuen-2000}.  Mims \cite{Mims-1972} has also noted that the presence of a frozen-core of nuclear spins around the electron spin can extend its coherence time.  This frozen core has also been observed Pr$^{3+}$-doped LaF$_3$ in optically-detected ESR experiments \cite{Shelby-1978,Wald-1992}, and in ruby \cite{Szabo-1990}.  Combined microwave and optical techniques have been used to analyze the barrier in fluorene-h10 doped with fluorene-d10 \cite{Verheij-1993}.

Coherent neutron scattering experiments have also been used to probe polarized nuclei close to paramagnetic impurities \cite{Stuhrmann-1997,Stuhrmann-2007}, yielding an estimate of about 1nm for the spin-diffusion barrier.  Neutron scattering could be an important tool in exploring the physics of the near-nuclei, given the large difference in the proton scattering cross-section of spin polarized neutrons, depending on whether the two spins are aligned or anti-aligned \cite{Abragam-1982}.  

\subsubsection*{Bulk spin diffusion of Zeeman and dipolar energy}

The signal observed in an NMR experiment is usually not from the spins closest to the electron spin.  In addition to being much smaller in number, the spins in and around the diffusion barrier are both frequency-shifted and broadened compared to the nuclear spins in the bulk.  The nuclear spins in the bulk only experience the presence of the electrons indirectly --- mediated by bulk spin diffusion as discussed above.

Since Bloembergen's original description of spin diffusion and its role in the role in mediating spin-lattice relaxation in ionic solids, the process has been studied extensively in a variety of materials.  It was soon understood that Zeeman and dipolar energy can be transported independently in the high field limit.  
Historically, attempts to measure the spin diffusion rate of Zeeman \cite{Leppelmeier-1968,Gates-1977} and dipolar energy \cite{Furman-1999} relied on the theoretical models that related the diffusion rate to observed relaxation times. There have also been a number of attempts to calculate the rate of spin diffusion in single
crystals, for both the Zeeman energy and the dipolar energy of the spin system using theoretical models \cite{Bloembergen-1949,Khutsishvili-1957,deGennes-1958,Redfield-1959,Lowe-1967,Borckmans-1968,Redfield-1968,Redfield-1969,Borckmans-1973,Greenbaum-2005,Kuzemsky-2006}, and classical simulation \cite{Tang-1992,Sodickson-1995}.

Spin diffusion provides a well-posed problem in the study of multi-body dynamics, as the Hamiltonian of the system is well known, and the nuclear spins are well isolated from other degrees of freedom in the crystal.  The study of the diffusion of the Zeeman and dipolar ordered states is essentially the study of the evolution of different initial states under the secular dipolar Hamiltonian. In particular the Zeeman ordered state consists of single spin population terms only, while the dipolar ordered state consists of correlated two spin states \cite{Cho-2003}. In a strong magnetic field these quantities are independently conserved, and have different diffusion rates and spin-lattice relaxation times.

During the spin diffusion process, the spins are in dynamical equilibrium under the action of the secular dipolar Hamiltonian, and the constants of the motion are only defined for the total spin system.  If the identity of individual spins can be distinguished, the spins are clearly seen to evolve in time as in the case where we write a position dependent phase on the spins. This is the essence of reciprocal space diffusion measurements.  Zhang and Cory used reciprocal space techniques to perform the first direct experimental measurement of spin diffusion in a crystal of CaF$_2$ \cite{Zhang-1998a}.  We recently extended this technique to directly measure the spin diffusion rate of dipolar energy \cite{Boutis-2004}.  While theoretical studies have suggested that the spin diffusion rates of Zeeman and dipolar order should approximately be the same, we found that dipolar diffusion was significantly faster than Zeeman diffusion (see Table 1).  

\begin{table}
\caption{Summary of the experimental results of the spin diffusion
rate of spin-spin energy, $D_{D}$, and Zeeman energy, $D_{Z}$ for
single crystal calcium fluoride. (from Ref.(62))}
\begin{tabular} {||c|c|c|c||} \hline Ref.\ \cite{Boutis-2004}
&[001]&[111]&$D_{001}/D_{111}$  \\ \hline
\hspace*{0.1in} $D^{||}_{D}$ \hspace*{0.1in} ($ \times 10^{-12}$cm$^{2}$/s) & 29 $\pm$ 3& 33 $\pm$ 4 & 0.88 $\pm$ 0.14 \\
\hspace*{0.1in} $D^{||}_{Z}$ \hspace*{0.1in} ($ \times 10^{-12}$cm$^{2}$/s) & 6.4 $\pm$ 0.9 & 4.4 $\pm$ 0.5 & 1.45 $\pm$ 0.26  \\
 \hline
Ref.\ \cite{Zhang-1998a}& [001] &[111]&$D_{001}/D_{111}$ \\
\hline
\hspace*{0.1in} $D^{||}_{Z}$ \hspace*{0.1in}  ($ \times 10^{-12}$cm$^{2}$/s) & 7.1 $\pm$ 0.5 & 5.3 $\pm$ 0.3 &1.34 $\pm$ 0.12
\\
\hline

Theoretical studies of $D_{Z}^{||}$& [001] & [111] &$D_{001}/D_{111}$ \\ \hline

Ref. \cite{Borckmans-1968} ($ \times 10^{-12}$cm$^{2}$/s) & 6.98 &  4.98 & 1.4\\
Ref. \cite{Redfield-1969}  ($\times 10^{-12}$cm$^{2}$/s) & 8.22  & 6.71 & 1.22 \\
Ref. \cite{Tang-1992}  ($\times 10^{-12}$cm$^{2}$/s) & 7.43& -- & -- \\ \hline

Theoretical studies of $D_{D}^{||}$& [001] & [111]& $D_{001}/D_{111}$ \\ \hline

Ref. \cite{Borckmans-1973}  ($\times 10^{-12}$cm$^{2}$/s) & 8.53 &  8.73 & 0.98 \\
Ref. \cite{Tang-1992}   ($\times 10^{-12}$cm$^{2}$/s) & 13.3& -- & -- \\ \hline

Ratio of $D_{D}$ to $D_{Z}$& [001] & [111] & \\ \hline
Ref.\ \cite{Boutis-2004} & $4.5\pm0.8$ & $7.5\pm1.3$& -- \\
Ref.\ \cite{Borckmans-1973}& 1.22 & 1.75 & --  \\
Ref.\ \cite{Tang-1992} & 1.79 & -- & -- \\ \hline
\end{tabular}
\label{tableofresults}
\end{table}

An examination of the physical processes leading to the diffusion shows that the dynamics can be quite different.  Figure 2 shows a simple illustrative model that suggests that the diffusion of dipolar order should be faster (and more complicated) than that of Zeeman order, due to the increase in the number of possible paths for the propogation of the dipolar ordered state.  The rapid diffusion of dipolar order is very likely a consequence of a constructive interference effect in the many-spin dynamics

Recently a direct real space measurement of spin diffusion was made using a magnetic resonance force microscope \cite{Eberhardt-2007}.  They extracted  a Zeeman diffusion coeffient of  $D_Z = (6.2 \pm 0.7) \times 10^{-12}$ cm$^2$/s, and estimated the dipolar diffusion rate to be $D_D = (11 \pm 11) \times 10^{-12}$ cm$^2$/s from their measurement, in agreement with earlier results.

\begin{figure}
\scalebox{0.6}{
\includegraphics{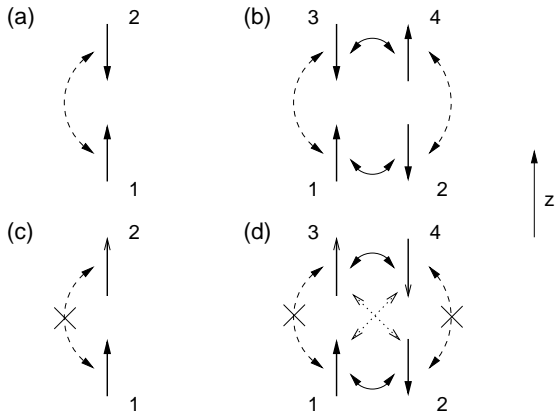}}
\caption{(a) For the diffusion of Zeeman order along the z direction spins 1 and 2 need to undergo a flip-flop.  (b) For the diffusion of dipolar order along the z direction, there are two possibilities, both spins 1 and 3, and spins 2 and 4 can flip, or spins 1 and 2, and spins 3 and 4 can flip.  There are thus two different paths to the same final state and interference effects may be observed. (c) If spins 1 and 2 are initially in the same state (both $\uparrow$ or both $\downarrow$),  no evolution takes place. (d) Even if states 3 and 4 are initially  $\uparrow$ and $\downarrow$, two different evolution paths are present.  Spins 1 and 2 can flip and spins 3 and 4 can flip or spins 1 and 4 and spins 2 and 3 can flip.  Thus dipolar diffusion dynamics are less easily quenched. (from Ref.(62))} 
\end{figure}

\subsubsection*{Recent experiments}

A number of recent experiments have re-emphasized the importance of the role of both the spin diffusion barrier and bulk spin diffusion in the DNP process.  
Michal and Tycko \cite{Michal-1998} observed the creation of optically pumped dipolar order in the $^{115}$In spins of indium phosphide.  They suggested that the low dipolar spin temperature they observed could be the produced indirectly by polarization transport through the field gradients of the trapped electrons at the optical pumping sites as proposed by Genack and Redfield \cite{Genack-1973,Genack-1975} or directly by optical pumping.  Tycko \cite{Tycko-1998} noted that nuclear-spin dipolar order can directly result from optical pumping if the hyperfine couplings are of the dipolar form.  Patel and Bowers \cite{Patel-2004} used multiple-quantum NMR techniques to show the creation of dipolar order in both gallium arsenide and indium phosphide, following optical pumping.

In a microwave-induced DNP experiment on a 40 mM frozen solution of 4-amino TEMPO (in a 40:60 water/glycerol mixture), we recently observed that the bulk proton dipolar reservoir is cooled to a spin temperature that is significantly lower than the Zeeman spin temperature \cite{Dementyev-2008a}.  In addition we were able to enhance the NMR signal by 50 \% by equilibrating the the temperatures of the nuclear Zeeman and dipolar reservoirs.  We believe that a Genack and Redfield mechanism is responsible for producing the low dipolar spin temperature in the vicinity of the electron spins.  Moreover, as we observed in our earlier spin diffusion measurements, dipolar spin diffusion is significantly faster than Zeeman spin diffusion \cite{Zhang-1998a,Boutis-2004}, and the bulk dipolar reservoir cools faster than the bulk Zeeman reservoir.  In principle, this process can be exploited to rapidly polarize the nuclear spins, by repeatedly cooling the dipolar system and transferring the polarization to the Zeeman reservoir.

The ability to increase the Zeeman magnetization via contact with the dipolar reservoir is exciting, as it should be possible to polarize a sample more rapidly by repeatedly cooling the dipolar reservoir and transferring this polarization to the Zeeman reservoir (see Figure 3).  Note that this transfer of order occurs in the bulk crystal, not just locally to the defect sites.  When the weak spin-locking field is applied the cross polarization between the Zeeman and dipolar systems occurs on a much faster timescale, since it does not require macroscopic transport of the polarization.  The transport is not limited by the small heat capacity of the dipolar reservoir.

\begin{figure}
\scalebox{0.5}{\includegraphics{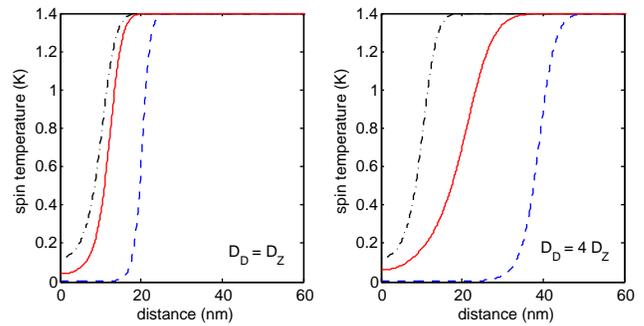}} \caption{(Color online) Zeeman (black dash-dot line) and dipolar (blue dashed line) spin temperatures  following 1 second of spin diffusion, obtained from simulations where (a) $D_D = D_Z$ and (b) $D_D = 4 D_Z$.  The inital spin temperatures at the left edge are 10 $\mu$K and 20 mK for the dipolar and Zeeman reservoirs respectively.  The solid red line indicates the final spin temperature following adiabatic transfer. (from Ref.(8))
\label{fig:sims}}
\end{figure}

Tycko has suggested that since semiconductor materials like indium phosphide can be hyperpolarized by optical pumping techniques, it might be possible to polarize organic or biological systems that are deposited on such substrates by polarization transfer processes \cite{Tycko-1998}.  In addition to spin diffusion from the optical pumping sites to the bulk, this process requires spin diffusion across the semiconductor/organic interface during the cross-polarization process.  Goehring and Michal recently observed that it was possible to transfer approximately 20 \% of the total nuclear spin polarization from micron sized InP particles to an organic layer on the surface \cite{Goehring-2003} using Hartmann-Hahn techniques \cite{Hartmann-1962}.

Polarization transport by spin diffusion across such a heterogeneous interface following DNP has also been observed.  Griffin and co-workers have demonstrated that the DNP-enhanced nuclear spin polarization of the protons of an aqueous solvent (containing the biradical TOTAPOL), could be transferred to the protons of nanocrystals of the peptide GNNQQNY, via spin diffusion \cite{vanderWel-2006}.  Though their nanocrystals were on the order of 100-200 nm, their model suggests that the nanocrystals upto 1 $\mu$m in diameter could be efficiently polarized using this technique.  In a related experiment, we found that microwave irradiation of the electrons (dangling bonds) of the amorphous surface layer of silicon microparticles (range 1-5 $\mu$m) produced a large dynamic nuclear polarization, which was eventually transferred to the crystalline core by spin diffusion \cite{Dementyev-2008b}.  X-ray diffraction revealed that this sample was approximately 80 \% amorphous and 20 \% crystalline.  Though the surface nuclei had a relatively short T$_1$ (on the order of minutes), the nuclear spins in the crystalline core had very long T$_1$s (on the order of a few hours), as the T$_1$ of the core is mediated by spin diffusion to the surface (T$_1$ $\approx$ R$^2$/D).  The long relaxation times makes these particles a candidate for tracer studies.

\subsubsection*{Challenges}
In the above discussion, we have ignored the details of the electron spin relaxation process, though the influence of the phonon bottleneck at low temperatures  is well known \cite{Abragam-1982}.  
As can be seen from the above discussion, the polarization of bulk nuclei following DNP irradiation is a complex, multi-step process.  Many DNP experiments are characterized by trial and error, rather than a first principles design.  Efforts to develop optimal control techniques for DNP are likely to be focused on a single step, until we can deal with the complexity of the many-body dynamics involved.  The models dealing with the spin-diffusion barrier are very approximate, as they ignore the discrete nature of the lattice and depend on a continuum transport model.  Bulk spin diffusion also remains an open problem in many-body spin dynamics.  The physics of these systems is rich and complex, and care should be taken when dealing with simple models.  The field, temperature, concentration of electron spins, and the nature of the nuclear spin system all strongly influence the DNP process and the particular experimental conditions can strongly determine the outcome.

\vspace*{0.1in}
\begin{center}
{\em Acknowledgements}  
\end{center}
C.R. thanks Gregory Boutis, HyungJoon Cho, David Cory, Anatoly Dementyev, Daniel Greenbaum and Jonathan Hodges for stimulating discussions.  This work was supported in part by the National Security Agency (NSA) under Army Research Office (ARO) contract number W911NF0510469, the NSF and DARPA DSO.
\bibliography{Bibliography}
\end{document}